\documentclass[aps,twocolumn,showpacs,preprintnumbers,amsmath,amssymb,floatfix]{revtex4}
\usepackage{graphicx}% Include figure files
\usepackage{dcolumn}% Align table columns on decimal point
\usepackage{bm}% bold math
\usepackage[german,american]{babel}

\begin{document}
%\preprint{APS/123-QED}

\title{Cation Transport in Polymer Electrolytes: A Microscopic Approach}

\author{A. Maitra}
 %\\altaffiliation{Physics Department, XYZ University.}%Lines break automatically or can be forced with \\
\author{A. Heuer}%
 %\\email{Second.Author@institution.edu}

\affiliation{
Westf\"{a}lische Wilhelms-Universit\"{a}t M\"{u}nster, Institut f\"{u}r Physikalische Chemie,
Corrensstr. 30, 48149 M\"{u}nster, Germany }
\affiliation{
NRW Graduate School of Chemistry, Corrensstr. 36, 48149 M\"{u}nster, Germany }

\date{\today}% It is always \today, today,

\begin{abstract}
A microscopic theory for cation diffusion in polymer electrolytes
is presented. Based on a thorough analysis of molecular dynamics
simulations on PEO with LiBF$_4$ the mechanisms of cation dynamics
are characterised. Cation jumps between polymer chains can be identified 
as {\it renewal} processes. This allows us to obtain an explicit expression
for the lithium ion diffusion constant $D_{Li}$ by invoking polymer specific 
properties such as the Rouse dynamics. This extends previous phenomenological and 
numerical approaches.
In particular, the chain length
dependence of $D_{Li}$ can be predicted and compared with
experimental data. This dependence can be fully understood without
referring to entanglement effects.
\end{abstract}

 %PACS, the Physics and Astronomy
\maketitle

Transport of cations in complex systems is of major relevance in
the field of disordered ion conductors. Specifically, polymer
electrolytes \cite{GraySPE,BruceVin93,RatShr88}, using lithium
salts, have been intensively studied experimentally and
theoretically due to their technological relevance. Free lithium
ions (Li$^+$), uncomplexed by the anions, are the desirable charge
carriers in electrolytic applications. A theoretical description
of ionic dynamics in terms of microscopic properties is  difficult
because the dynamics of the cations and the polymer segments occur
on the same time scale \cite{RatnerMRS,BoroMDLiBF4}. In contrast,
the dynamics of ions in inorganic systems can be characterized by
ionic hops between permanent sites, supplied by the immobile
network \cite{LamHeuprl}.

\begin{figure}
  %\vspace{4pt}
  \centering
  \includegraphics[width=0.6\linewidth]{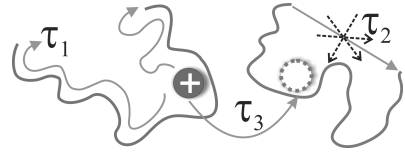}
  \caption{\label{fig1}Time scales: $\tau_1$ is the time scale for intra-chain ionic motion (M1),
  $\tau_2$ is the relaxation time of the polymer chain (related to $\tau_{R}$: see text for details)(M2),
  and $\tau_3$ is the waiting time of an ion between two inter-chain jumps (M3).}
\end{figure}

The phenomenological dynamic bond-percolation model (DBP) \cite{DynDisTrans94} 
considers that the long-range ion transport is enabled by {\it renewal} events which 
lift the blockages in the ionic pathways. Since the dynamics after a renewal 
event is statistically uncorrelated to its past, the resulting cationic diffusion constant, 
$D_{Li}$, is determined by  $a^2/6\tau_{ren}$ where $a^2$ and $\tau_{ren}$ denote 
the typical mean square displacement (MSD) and the time period between 
two consecutive renewal events, respectively. In the DBP 
the renewal process is attributed to a local structural relaxation process  
governed by the polymer dynamics. 
Another fruitful approach to understand the mechanisms of cation
dynamics is by means of molecular dynamics (MD) simulations \cite{Allen,Plathe,Neyertz95,BoroLiIStatic,Dynlichlo}.
Cationic dynamics can be divided
into three important mechanisms: (M1) motion along a chain
("intra-chain"), (M2) motion together with chain segments, using
the chain as a vehicle ("segmental"), and (M3) jumps between
different chains ("inter-chain"); see also \cite{BoroMDLiTFSI}.
This is sketched in Fig.\ref{fig1}.
Based on the insight from  
MD simulations, Borodin and Smith \cite{BoroMDLiTFSI,BoroIoniLiq} have 
recently formulated a microscopic transport model. Employing appropriately defined Monte-Carlo
moves and implicitly using the concept of renewal process the Li$^+$ dynamics 
has been reproduced. Among other things, they have quantified the relevance of the variety of cation 
transport mechanisms that contribute to the macroscopic cation diffusivity $D_{Li}$. 

Our methodology is founded on both the approaches. First, we 
express $a^2$ in terms of (M1) and (M2) by exploiting the fact 
that the Li$^{+}$ dynamics is strongly correlated to the
polymer segmental dynamics, which in turn can be separated into 
statistically uncorrelated center-of-mass (c.o.m) dynamics of the polymer chain 
(zeroth order Rouse mode) and its internal dynamics (higher order Rouse modes) \cite{DoiEd}. 
This implies, for the Li$^+$ ions, $D_{Li} = D_{c.o.m} + D_M(\tau_1, \tau_2, \tau_3)$.
The time scales $\tau_1$, $\tau_2$ and $\tau_3$
characterize each of the mechanisms (M1), (M2) and (M3), respectively.
Extending Ref. \cite{BoroMDLiTFSI}, 
we derive analytical formulas in terms of 
$\tau_i$ with correlations between (M1) and (M2) being taken into account. 
The time scales are extracted from long MD simulations 
of a model polymer electrolyte system. Second, we explicitly show that (M3) 
can be identified as a renewal process so that, indeed, 
a prediction of the long-time behavior, i.e. $D_{Li}$, becomes possible. As a central feature we 
obtain the $N$-dependence ($N$: number of monomers per chain) of 
the $\tau_i$ and thus of $D_{Li}$. Two different $N$-regimes are obtained 
in agreement with experimental data. 

Atomistic NVT MD simulations are performed for the system
poly(ethylene) oxide  (PEO) and LiBF$_4$ with a concentration of
EO:Li=20:1 (EO: ether oxygen). The two body effective polarizable
potential employed is described in Ref. \cite{BoroMDLiBF4}. We
have simulated this system for different chain lengths ($N=24$ and
$N=48$) and at different temperatures (400 K $\le T \le $ 450 K).
The respective densities have been chosen to set the average pressure to 
values of the order of 1 MPa.
This letter discusses the results for the $N=48$ system at $T
= 450$ K unless specified otherwise.

\begin{figure}
  \includegraphics[width=0.7\linewidth]{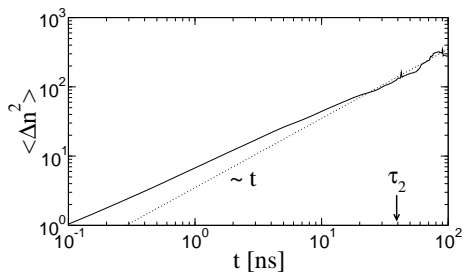}
  \caption{\label{fig2}Average square variation of the average oxygen index of
  one chain to which a Li$^{+}$ is associated during time interval $t$.}
  %\vspace{-5pt}
\end{figure}

As has been observed before
\cite{BoroMDLiBF4,BoroLiIStatic,Neyertz95,Plathe} a Li$^+$ ion
is, most of the time, coordinated to a single polymer chain
through EO atoms, interrupted by infrequent transitions between
different chains.
A Li$^+$ ion which is bound to a polymer chain is coordinated to a
few ($\approx 5$) and mostly contiguous oxygen atoms. After serially
indexing the oxygen atoms of a chain in succession we mark the
position of the Li$^+$ at the chain by the average index, $n(t)$, of
the enumerated oxygen atoms that belong to its coordination
sphere. To elucidate (M1), we determine the average-square variation
$\langle \Delta n(t)^2 \rangle$ of the average oxygen index under
the constraint that the Li$^+$ ion is attached to the same chain
during the time interval of length $t$. The result is shown in
Fig.\ref{fig2}. To a good approximation one observes diffusive
dynamics $\langle \Delta n(t)^2 \rangle = 2D_1 t $ where $D_1$ is
the intra-chain ionic diffusivity. To account for the slight
deviations from linear behavior we choose $D_1 = \langle \Delta
n(\tau_2)^2 \rangle/ 2\tau_2$ ($\tau_2$ defined in the next paragraph).
For later purposes we define
\begin{equation}
\tau_1 = (N-1)^2  / \pi^2 D_1
\end{equation}
where $\tau_1$ is a measure of the time it takes for the lithium ion to
diffuse from one end to the other end of the polymer chain. Here we
obtain $\tau_1 \approx 150$ ns (with $D_1$ from Fig.\ref{fig2}).

\begin{figure}
  %\vspace{4pt}
  \includegraphics[width=8.6cm]{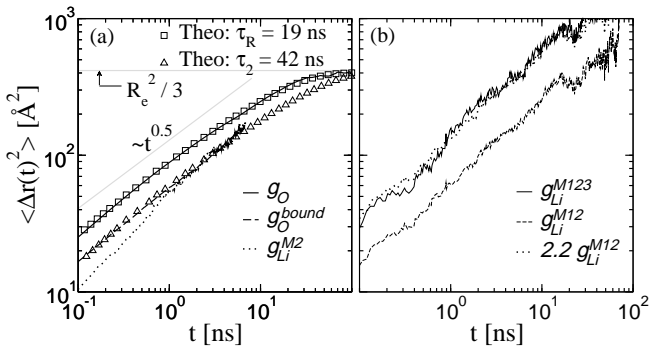}
  \caption{\label{fig3}(a) MSD of (i) all oxygen atoms (solid), (ii) oxygen atoms
which are bound to one $Li^{+}$ during time duration $t$ (dash), (iii) $Li^{+}$ which  changed the  average index of its oxygen neighbors by at most 1 (dot). Also shown are the Rouse  predictions  with $\tau_{R}$=19 ns ($\scriptstyle{\square}$) and $\tau_{2}$=42 ns ($\scriptstyle{\triangle}$).
  (b) MSD of Li$^{+}$ under different constraints (see text). }
  %\vspace{-5pt}
\end{figure}

To characterize (M2) we first analyze the polymer dynamics. In
Fig.\ref{fig3}(a) we display the MSD $g_O(t)$ for an average
oxygen atom (i.e. all oxygen atoms were considered for analysis
irrespective of the presence or absence of Li$^+$ near an oxygen
atom), characterizing the dynamics of the polymer segments.
According to our general procedure all MSD-functions are computed
relative to the polymer c.o.m. It exhibits a 
Rouse-like behavior \cite{Rou53,DoiEd} for short times $g_O(t) \propto t^\alpha$ with
$\alpha \approx 0.6$, saturating at $g_O(t) \approx R_e^2/3$ where
$R_e^2$ is the mean square end-to-end distance of the polymer
. We have included the theoretical Rouse
prediction, obtained via numerical summation
\begin{equation}\label{rouse}
g(t/\tau_R) = \frac{2R_e^2}{\pi^2} \sum_{p=1}^{N-1} \frac{1 -
\exp(-\frac{p^2t}{\tau_{R}})}{p^2}
\end{equation}
where $\tau_R$ and $p$ denote Rouse time and mode number, respectively, 
and the sum  
is calculated over the $N-1$ eigenmodes.
It yields a reasonable description
of the observed MSD, using $\tau_R = 19$ ns; see also
\cite{Kreer2001}. Qualitatively, it is expected that the oxygen
atoms which are temporarily bound to a Li$^+$ ion ought to be
somewhat slower due to the decrease in the local degrees of freedom.
This was checked by calculating $g_{O}^{bound}(t)$ for those oxygen
atoms which, during the whole time interval of length $t$, are bound
to one Li$^+$ ion. Indeed, $g_{O}^{bound}(t)$ is also consistent
with the Rouse prediction, using a longer Rouse time $\tau_2 = 42$
ns. Naturally, $\tau_2/\tau_R >1$ reflects the immobilization of the
polymer segments due to the ions. Note that for longer $t$ less
oxygen atoms contribute to this curve so that the statistics gets
worse.

Switching to the Li$^+$ dynamics, we first calculate the MSD
$g_{Li}^{M2}$(t) of Li$^+$ ions for which $|n(t)-n(0)| \leq 1$; see Fig.\ref{fig3}(a). For $t > 2$ ns this curve
is close to $g_{O}^{bound}(t)$. Thus, we can conclude that the
Li$^+$ motion strictly follows the oxygen dynamics in the absence
of (M1). In other words, the cations and the polymer segments
exhibit coupled dynamics.
Additionally, we find 
that for shorter times the Li$^+$ ions are slower than the
corresponding oxygen atoms.

In the absence of ion jump events between chains (M3) one would
have $D_M = 0$.  We have identified the jumps from a microscopic
analysis of the trajectories. On average after $\tau_3 = 110$ ns a
Li$^+$ ion jumps between two chains. To characterize the effect of
these jumps we have first determined $g_{Li}^{M123}(t)$ which is
the MSD of a Li$^+$ ion between times $[t_0-t,t_0+t]$ if at time
$t_0$ a jump happens and during the intervals $[t_0-t,t_0]$ and
$[t_0,t_0+t]$ the ion stays with the same chain, respectively. The
MSD is averaged over all jumps (Fig.\ref{fig3}(b)).
Furthermore we have determined the MSD $g_{Li,\pm}^{M12}(t)$ during the
intervals $[t_0-t,t_0]$ or $[t_0,t_0+t]$,  i.e. just before or after
an inter-chain jump, respectively. 
For symmetry reasons one has $g^{M12}_{Li,+}(t) = g^{M12}_{Li,-}(t) \equiv g^{M12}_{Li}(t)$. 
If, additionally, the inter-chain jump at $t_0$ serves as a renewal process, 
the statistical independence of cationic dynamics before and after the jump 
requires $g_{Li}^{M123} (t) = g^{M12}_{Li,+}(t)  +g^{M12}_{Li,-}(t) = 2g^{M12}_{Li}(t)$.
In case of correlations a smaller factor is
expected compared to 2. As exhibited in Fig.\ref{fig3}b
this relation is indeed found with minor deviations (prefactor 2.2
instead of 2.0), thus validating the fact that 
inter-chain transitions can be regarded as renewal processes.

In the following, we derive an explicit expression $D_M =
D_M(\tau_1,\tau_2,\tau_3)$. Based on the observed correlations
between a Li$^+$ ion and the polymer dynamics, $g_{Li}^{M12}(t)$ can be
described by taking into account the Rouse dynamics (M2) plus the
additional intra-chain diffusion (M1). Formally, one can write
\begin{equation}
g_{Li}^{M12}(t) = \langle (\vec{r}_j(t) - \vec{r}_i(0) )^2
\rangle_{M12}
\end{equation}
The average is over the probability density that the initial
oxygen index to which an ion is linked is $i$ and at a time $t$
later is $j$ and the distribution of monomer displacements as
predicted in the Rouse theory. From Ref.\cite{DoiEd} one obtains,
using $\langle \cos^2(p\pi (j-0.5)/N) \rangle_{M1} = 1/2$ and
$\langle \cos(p \pi (i+j-1)/N) \rangle_{M1} = 0$,
\begin{equation}\label{g2}
g_{Li}^{M12}(t) = \frac{2R_e^2}{\pi^2} \sum_{p=1}^{N-1}
\frac{1 - \langle \cos \frac {p\pi (i-j)}{N}
\rangle_{M1}\exp(-\frac{p^2t}{\tau_2})}{p^2}.
\end{equation}
Assuming Gaussian dynamics for (M1), i.e.  $\langle \Delta n^2 \rangle \propto t$
(correction to the deviation as observed in
Fig.\ref{fig2} can be easily implemented but are not of relevance
here), one finds $\langle \cos(p \pi (i-j)/N) \rangle_{M1} =
\exp(-p^2 t/\tau_1)$. Eq. \ref{g2} simplifies to (using
Eq.\ref{rouse})
\begin{equation}\label{g2_1}
g_{Li}^{M12}(t) = g(t/\tau_{12})
\end{equation}
with  $1/\tau_{12} = 1/\tau_1 + 1/\tau_2$. Thus, the dynamical
effects of (M1) and (M2) appear through the resulting relaxation
rate $1/\tau_{12}$. Finally, using the renewal property one can
write $D_M = a_M^2/6\tau_3$ explicitly as
\begin{equation}
\label{renewal}  D_M =  \langle g_{Li}^{M12}(\tau) \rangle_{M3}/6
\tau_3
\end{equation}
where a$_M^2$ corresponds to the average MSD between successive inter-chain hopping
events due to (M1) and (M2).
The average is taken over the distribution of time intervals
between these events.
For the numerical analysis (see below) we take a simple exponential 
distribution.

Approximate analytical expressions can be obtained by converting the
sum in Eq.\ref{g2} into an integral from 0 to $\infty$ \cite{DoiEd}.
Then one obtains $g_{Li}^{M12} = 2R_e^2\pi^{-3/2}\sqrt{t/\tau_{12}}$
for $t \ll \tau_{12}$ and $g_{Li}^{M12} = R_e^2/3$ for $t \gg
\tau_{12}$, respectively. Inserting these results into
Eq.\ref{renewal} gives
\begin{subequations}\label{D_m}
\begin{align}
D_{M} & =\frac{R_e^2}{6\pi}\sqrt{\frac{1}{\tau_3\tau_{12}}}& \mbox{if $\tau_3 \ll \tau_{12}$}\label{1} \\
& =\frac{R_e^2}{18\tau_3}& \mbox{if $\tau_3 \gg \tau_{12}$}\label{2}
\end{align}
\end{subequations}
The scaling of $D_M$ with $\tau_3$ in Eq.\eqref{1} is consistent
with the numerical results, obtained in Ref.\cite{BoroMDLiTFSI}.
Eq.\eqref{1} holds for long chains and takes into
consideration the implicit correlations of (M1) and (M2).  By neglecting
these correlations, as done in Ref.\cite{BoroMDLiTFSI}, the term
$\sqrt{1/\tau_{12}} = \sqrt{1/\tau_1 + 1/\tau_{2}}$ would instead
become $\sqrt{1/\tau_1} + \sqrt{1/\tau_{2}}$. This would largely
overestimate $D_M$ (for instance, in PEO/LiBF$_4$ by 35 \%) and
the contribution of (M1) for the case $\tau_{3} \ll \tau_{12}$.

In contrast to the DBP-model we obtain $D_M \propto
1/\tau_{ren}$ only for short chains. The main conclusion,
however, that the Li$^+$ diffusion has the same temperature dependence
as the inverse Rouse time and thus as the polymer dynamics, remains
valid because all the three time scales $\tau_i$ have a similar temperature
dependence (data not shown).

In Tab.I we have compiled $\tau_1$, $\tau_2$, $\tau_3$, obtained from our
simulations for $N=24$ and $N=48$. Of major importance are their
scaling properties with $N$. One expects $\tau_1,\tau_2 \propto N^2$
and $\tau_3 \propto N^0$. Considering the appropriate number
of eigenmodes, the predictions for $N=24$, based on $N=48$, are given
in parenthesis. The agreement is convincing.
\begin{table}
  \begin{center}
     \begin{tabular}[c]{|c|c|c|c|} \hline
       $N$  & $\tau_1$ [ns] & $\tau_2$ [ns] & $\tau_3$ [ns]  \tabularnewline[1mm] \hline
       48 & 150   & 42  & 110   \tabularnewline[1mm] \hline
       24 &  34 (36)    &  10 (10) & 90 (110)  \tabularnewline[1mm] \hline
     \end{tabular}
   \caption{\label{tabsummary} The relevant time scales $\tau_i$ for $N=48$ and $N=24$ as well as a check of the
   scaling relations.}
   \vspace{-5pt}
  \end{center}
\end{table}
By setting $\tau_1 \rightarrow \infty$ one can estimate the contribution of
(M1) to $D_M$ in the limit of long chains ($\tau_3 \ll
\tau_{12}$). It is as small as 12\% for the present system.

Naively one might expect that $a_{M}^2 \propto N^0$ and thus $D_M
\propto N^0$ for all $N$, see e.g. \cite{ShiVin93}, because
(M1)-(M3) are related to local motions. Using the scaling results
for the $\tau_i$ one obtains, however, $D_M \propto N^0$ only for
long chains $(\tau_3 \ll \tau_{12})$ whereas $D_M \propto N$ for
short chains. The reason is that for short chains  $a_{M}^2$ is
limited by the end-to-end distance of the polymer which brings an
$N$-dependence.

\begin{figure}
  \includegraphics[width=8.6cm]{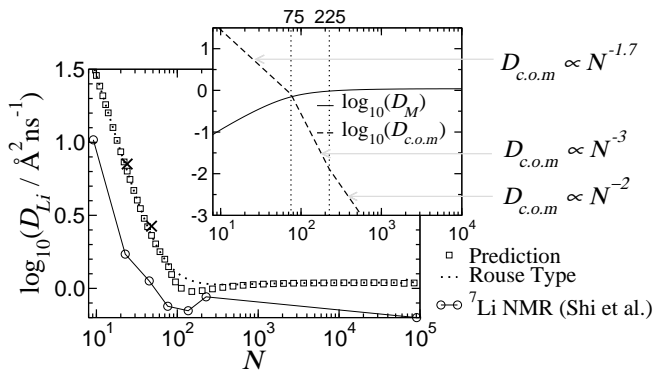}
  \caption{\label{fig4}Self diffusivity of Li$^+$, $D_{Li}=D_{M}+D_{c.o.m}$ with D{$_M$} from Eq.\ref{renewal}  ($\scriptstyle{\square}$).
  The dotted line shows the prediction for a pure Rouse-type motion. Experimental\cite{ShiVin93}
  $D_{Li}$ for PEO:LiCF$_3$SO$_3$ with EO:Li=20:1 at T=342 K is included ($\circ$). The inset shows  $D_{M}$ and $D_{c.o.m}$ individually.
 $D_{Li}$ values obtained from simulations are marked by crosses.}
 %\vspace{-5pt}
\end{figure}

In the present case, the
crossover in $D_M$, i.e. $\tau_3 \approx \tau_{12}$, occurs for $N \approx 100$
(see inset of Fig.\ref{fig4}). It has been speculated
\cite{ShiVin93} that the emergence of the entanglement regime ($N
\approx75 $ \cite{ShiVin93,ApFl1993}) leads to the crossover with an
accompanying change of cation conduction process from polymer c.o.m
dominated motion to percolation type transport (i.e. (M3)). However,
we find, it is a coincidence that the entanglement length is similar
to the crossover length. Thus, the crossover to $D_M \propto N^0$
is physically unrelated to entanglement effects.

We can predict the $N$-dependence of $D_{Li}(N)$ for
a large range of $N$ values by combining the empirical $N$-scaling
\cite{Meerwall98,Hayamizu1} (see inset Fig.\ref{fig4}) of
$D_{c.o.m}(N)$ with $D_{M}(N)$. The Rouse scaling ($D_{c.o.m}
\propto N^{-1}$) is known to be violated and substantially higher exponents
($\ge 1.5$) have been reported \cite{PSrev04,Meerwall98,Hayamizu1}.

To determine $D_{M}(N)$ we have used Eq.\ref{g2_1}
and \ref{renewal} by explicitly calculating the sum in Eq.\ref{g2}.
Fig.4 displays the predicted $D_{Li}(N)$ along with the experimental
data from Ref.\cite{ShiVin93}. In agreement, both indicate a
transition to a $N$-independent regime beyond $N \approx 100$.
Further included in Fig.\ref{fig4} is an estimation of $D_{Li}$
under the assumption that no entanglement effects are present for
$D_{c.o.m}$ by simply extending the scaling in the Rouse-regime to
all $N$.  As expected, the large-N behavior does not depend on the
specific form of $D_{c.o.m}$. However, one might speculate that the
appearance of the minimum in $D_M(N)$ is indicative of entanglement
effects.

Interestingly, within the present approach the value of $\tau_3$ can
be estimated from experimental data.  For very large $N=N_L$ one
has (see Eq.\eqref{1}) $\tau_3 = (N_sb^2/(6 \pi
D_{Li}|_{N=N_L}))^2/(\tau_2|_{N=N_S})$ where $b$ is the statistical
segment length of 
the polymer, $\tau_2/\tau_R \approx 2.2$ and $\tau_R|_{N=N_S}$ can
be estimated from $D_{c.o.m}\approx D_{Li}|_{N=N_S}$ (at some small
$N=N_S$) using the Rouse prediction. For the experimental data in
Fig.\ref{fig4} this yields $\tau_3 \approx 80$ ns. Via NMR experiments \cite{Hayamizu1}
local relaxation processes may be probed to extract information about $\tau_1$.

It is likely that the identified motional mechanisms are generally
applicable to understand the ionic dynamics in different polymer electrolytes.
Furthermore we checked that the presence of a
small fraction of ions which are, temporarily, not bound to a polymer
does not change the value of $D_M$ by more than 10\%. Of course,
corrections will become relevant for larger ion concentration due to
their mutual interaction. Note that our approach averages over the different
structural realizations and thus include, e.g., temporary complexation 
of an ion by two polymer chains. 

 In summary, we have elucidated ion dynamics in
polymer electrolytes by extracting microscopic properties from
simulation and expressing them in analytical terms. This extends
previous phenomenological approaches like the dynamic
bond-percolation model, by assigning its key concept, i.e. the
presence of {\it renewal} processes, a specific microscopic
interpretation.  For long chains the scaling relation $D_{Li}
\propto 1/\sqrt{\tau_{12} \cdot \tau_3}$ is obtained which goes
beyond the expressions from the DBP model. In any event, for the
transport of lithium ions fast transitions of ions between
different chains are vital. Since the expression
$D_{Li}(\tau_1,\tau_2,\tau_3)$ is now available one can estimate the
possible range of ionic mobilities for linear chain polymer electrolytes for all
$N$.

We would like to thank M. Vogel for important discussions and a
critical reading of the manuscript and Y. Aihara, J. Baschnagel, O. Borodin, K.
Hayamizu, and M. Ratner for helpful correspondence.

%\bibliography{reference3,notes}% Produces the bibliography via BibTeX.

\begin{thebibliography}{22}
\expandafter\ifx\csname natexlab\endcsname\relax\def\natexlab#1{#1}\fi
\expandafter\ifx\csname bibnamefont\endcsname\relax
  \def\bibnamefont#1{#1}\fi
\expandafter\ifx\csname bibfnamefont\endcsname\relax
  \def\bibfnamefont#1{#1}\fi
\expandafter\ifx\csname citenamefont\endcsname\relax
  \def\citenamefont#1{#1}\fi
\expandafter\ifx\csname url\endcsname\relax
  \def\url#1{\texttt{#1}}\fi
\expandafter\ifx\csname urlprefix\endcsname\relax\def\urlprefix{URL }\fi
\providecommand{\bibinfo}[2]{#2}
\providecommand{\eprint}[2][]{\url{#2}}

\bibitem[{\citenamefont{Gray}(1991)}]{GraySPE}
\bibinfo{author}{\bibfnamefont{F.~M.} \bibnamefont{Gray}},
  \emph{\bibinfo{title}{{Solid Polymer Electrolytes}}}
  (\bibinfo{publisher}{Wiley-VCH, New York}, \bibinfo{year}{1991}).

\bibitem[{\citenamefont{Bruce and Vincent}(1993)}]{BruceVin93}
\bibinfo{author}{\bibfnamefont{P.~G.} \bibnamefont{Bruce}} \bibnamefont{and}
  \bibinfo{author}{\bibfnamefont{C.~A.} \bibnamefont{Vincent}},
  \bibinfo{journal}{J. Chem. Soc. Faraday Trans.}
  \textbf{\bibinfo{volume}{89}}, \bibinfo{pages}{3187} (\bibinfo{year}{1993}).

\bibitem[{\citenamefont{Ratner and Shriver}(1988)}]{RatShr88}
\bibinfo{author}{\bibfnamefont{M.~A.} \bibnamefont{Ratner}} \bibnamefont{and}
  \bibinfo{author}{\bibfnamefont{D.~F.} \bibnamefont{Shriver}},
  \bibinfo{journal}{Chem. Rev.} \textbf{\bibinfo{volume}{88}},
  \bibinfo{pages}{109} (\bibinfo{year}{1988}).

\bibitem[{\citenamefont{Ratner et~al.}(2000)\citenamefont{Ratner, Johansson,
  and Shriver}}]{RatnerMRS}
\bibinfo{author}{\bibfnamefont{M.}~\bibnamefont{Ratner}},
  \bibinfo{author}{\bibfnamefont{P.}~\bibnamefont{Johansson}},
  \bibnamefont{and} \bibinfo{author}{\bibfnamefont{D.}~\bibnamefont{Shriver}},
  \bibinfo{journal}{MRS Bulletin} \textbf{\bibinfo{volume}{25}},
  \bibinfo{pages}{31} (\bibinfo{year}{2000}).

\bibitem[{\citenamefont{Borodin et~al.}(2003)\citenamefont{Borodin, Smith, and
  Douglas}}]{BoroMDLiBF4}
\bibinfo{author}{\bibfnamefont{O.}~\bibnamefont{Borodin}},
  \bibinfo{author}{\bibfnamefont{G.~D.} \bibnamefont{Smith}}, \bibnamefont{and}
  \bibinfo{author}{\bibfnamefont{R.}~\bibnamefont{Douglas}},
  \bibinfo{journal}{J. Phys. Chem. B} \textbf{\bibinfo{volume}{107}},
  \bibinfo{pages}{6824} (\bibinfo{year}{2003}).

\bibitem[{\citenamefont{Lammert et~al.}(2003)\citenamefont{Lammert, Kunow, and
  Heuer}}]{LamHeuprl}
\bibinfo{author}{\bibfnamefont{H.}~\bibnamefont{Lammert}},
  \bibinfo{author}{\bibfnamefont{M.}~\bibnamefont{Kunow}}, \bibnamefont{and}
  \bibinfo{author}{\bibfnamefont{A.}~\bibnamefont{Heuer}},
  \bibinfo{journal}{Phys. Rev. Lett.} \textbf{\bibinfo{volume}{90}},
  \bibinfo{pages}{215901} (\bibinfo{year}{2003}).

\bibitem[{\citenamefont{Nitzan and Ratner}(1994)}]{DynDisTrans94}
\bibinfo{author}{\bibfnamefont{A.}~\bibnamefont{Nitzan}} \bibnamefont{and}
  \bibinfo{author}{\bibfnamefont{M.~A.} \bibnamefont{Ratner}},
  \bibinfo{journal}{J. Phys. Chem.} \textbf{\bibinfo{volume}{98}},
  \bibinfo{pages}{1765} (\bibinfo{year}{1994}).

\bibitem[{\citenamefont{Allen and Tildesley}(2004)}]{Allen}
\bibinfo{author}{\bibfnamefont{M.~P.} \bibnamefont{Allen}} \bibnamefont{and}
  \bibinfo{author}{\bibfnamefont{D.~J.} \bibnamefont{Tildesley}},
  \emph{\bibinfo{title}{{Computer Simulation of Liquids}}}
  (\bibinfo{publisher}{Clarendon, Oxford}, \bibinfo{year}{2004}).

\bibitem[{\citenamefont{M\"{u}ller-Plathe and van Gunsteren}(1995)}]{Plathe}
\bibinfo{author}{\bibfnamefont{F.}~\bibnamefont{M\"{u}ller-Plathe}}
  \bibnamefont{and} \bibinfo{author}{\bibfnamefont{W.}~\bibnamefont{van
  Gunsteren}}, \bibinfo{journal}{J. Chem. Phys.}
  \textbf{\bibinfo{volume}{103}}, \bibinfo{pages}{4745} (\bibinfo{year}{1995}).

\bibitem[{\citenamefont{Neyertz and Brown}(1996)}]{Neyertz95}
\bibinfo{author}{\bibfnamefont{S.}~\bibnamefont{Neyertz}} \bibnamefont{and}
  \bibinfo{author}{\bibfnamefont{D.}~\bibnamefont{Brown}}, \bibinfo{journal}{J.
  Chem. Phys.} \textbf{\bibinfo{volume}{104}}, \bibinfo{pages}{3797}
  (\bibinfo{year}{1996}).

\bibitem[{\citenamefont{Borodin and Smith}(1998)}]{BoroLiIStatic}
\bibinfo{author}{\bibfnamefont{O.}~\bibnamefont{Borodin}} \bibnamefont{and}
  \bibinfo{author}{\bibfnamefont{G.~D.} \bibnamefont{Smith}},
  \bibinfo{journal}{Macromolecules} \textbf{\bibinfo{volume}{31}},
  \bibinfo{pages}{8396} (\bibinfo{year}{1998}).

\bibitem[{\citenamefont{Siqueira and Ribeiro}(2006)}]{Dynlichlo}
\bibinfo{author}{\bibfnamefont{L.~J.~A.} \bibnamefont{Siqueira}}
  \bibnamefont{and} \bibinfo{author}{\bibfnamefont{M.}~\bibnamefont{Ribeiro}},
  \bibinfo{journal}{J. Chem. Phys.} \textbf{\bibinfo{volume}{125}},
  \bibinfo{pages}{214903} (\bibinfo{year}{2006}).

\bibitem[{\citenamefont{Borodin and Smith}(2006)}]{BoroMDLiTFSI}
\bibinfo{author}{\bibfnamefont{O.}~\bibnamefont{Borodin}} \bibnamefont{and}
  \bibinfo{author}{\bibfnamefont{G.~D.} \bibnamefont{Smith}},
  \bibinfo{journal}{Macromolecules} \textbf{\bibinfo{volume}{39}},
  \bibinfo{pages}{1620} (\bibinfo{year}{2006}).

\bibitem[{\citenamefont{Borodin et~al.}(2006)\citenamefont{Borodin, Smith,
  Geiculescu, Creager, Hallac, and DesMarteau}}]{BoroIoniLiq}
\bibinfo{author}{\bibfnamefont{O.}~\bibnamefont{Borodin}},
  \bibinfo{author}{\bibfnamefont{G.~D.} \bibnamefont{Smith}},
  \bibinfo{author}{\bibfnamefont{O.}~\bibnamefont{Geiculescu}},
  \bibinfo{author}{\bibfnamefont{S.}~\bibnamefont{Creager}},
  \bibinfo{author}{\bibfnamefont{B.}~\bibnamefont{Hallac}}, \bibnamefont{and}
  \bibinfo{author}{\bibfnamefont{D.}~\bibnamefont{DesMarteau}},
  \bibinfo{journal}{J. Phys. Chem. B} \textbf{\bibinfo{volume}{110}},
  \bibinfo{pages}{24266} (\bibinfo{year}{2006}).

\bibitem[{\citenamefont{Doi and Edwards}(2003)}]{DoiEd}
\bibinfo{author}{\bibfnamefont{M.}~\bibnamefont{Doi}} \bibnamefont{and}
  \bibinfo{author}{\bibfnamefont{S.}~\bibnamefont{Edwards}},
  \emph{\bibinfo{title}{{The Theory of Polymer Dynamics}}}
  (\bibinfo{publisher}{Oxford Science Publications}, \bibinfo{year}{2003}).

\bibitem[{\citenamefont{Rouse}(1953)}]{Rou53}
\bibinfo{author}{\bibfnamefont{P.~E.} \bibnamefont{Rouse}},
  \bibinfo{journal}{J. Chem. Phys.} \textbf{\bibinfo{volume}{21}},
  \bibinfo{pages}{1272} (\bibinfo{year}{1953}).

\bibitem[{\citenamefont{Kreer et~al.}(2001)\citenamefont{Kreer, Baschnagel,
  M\"{u}ller, and K.Binder}}]{Kreer2001}
\bibinfo{author}{\bibfnamefont{T.}~\bibnamefont{Kreer}},
  \bibinfo{author}{\bibfnamefont{J.}~\bibnamefont{Baschnagel}},
  \bibinfo{author}{\bibfnamefont{M.}~\bibnamefont{M\"{u}ller}},
  \bibnamefont{and} \bibinfo{author}{\bibnamefont{K.Binder}},
  \bibinfo{journal}{Macromolecules} \textbf{\bibinfo{volume}{34}},
  \bibinfo{pages}{1105} (\bibinfo{year}{2001}).

\bibitem[{\citenamefont{Shi and Vincent}(1993)}]{ShiVin93}
\bibinfo{author}{\bibfnamefont{J.}~\bibnamefont{Shi}} \bibnamefont{and}
  \bibinfo{author}{\bibfnamefont{C.~A.} \bibnamefont{Vincent}},
  \bibinfo{journal}{Solid State Ionics} \textbf{\bibinfo{volume}{60}},
  \bibinfo{pages}{11} (\bibinfo{year}{1993}).

\bibitem[{\citenamefont{Appel and Fleischer}(1993)}]{ApFl1993}
\bibinfo{author}{\bibfnamefont{M.}~\bibnamefont{Appel}} \bibnamefont{and}
  \bibinfo{author}{\bibfnamefont{G.}~\bibnamefont{Fleischer}},
  \bibinfo{journal}{Macromolecules} \textbf{\bibinfo{volume}{26}},
  \bibinfo{pages}{5520} (\bibinfo{year}{1993}).

\bibitem[{\citenamefont{von Meerwall et~al.}(1998)\citenamefont{von Meerwall,
  Beckman, Jang, and Mattice}}]{Meerwall98}
\bibinfo{author}{\bibfnamefont{E.}~\bibnamefont{von Meerwall}},
  \bibinfo{author}{\bibfnamefont{S.}~\bibnamefont{Beckman}},
  \bibinfo{author}{\bibfnamefont{J.}~\bibnamefont{Jang}}, \bibnamefont{and}
  \bibinfo{author}{\bibfnamefont{L.}~\bibnamefont{Mattice}},
  \bibinfo{journal}{J. Chem. Phys.} \textbf{\bibinfo{volume}{108}},
  \bibinfo{pages}{4299} (\bibinfo{year}{1998}).

\bibitem[{\citenamefont{Hayamizu et~al.}(2002)\citenamefont{Hayamizu, Akiba,
  Bando, and Aihara}}]{Hayamizu1}
\bibinfo{author}{\bibfnamefont{K.}~\bibnamefont{Hayamizu}},
  \bibinfo{author}{\bibfnamefont{E.}~\bibnamefont{Akiba}},
  \bibinfo{author}{\bibfnamefont{T.}~\bibnamefont{Bando}}, \bibnamefont{and}
  \bibinfo{author}{\bibfnamefont{Y.}~\bibnamefont{Aihara}},
  \bibinfo{journal}{J. Chem. Phys.} \textbf{\bibinfo{volume}{117}},
  \bibinfo{pages}{5929} (\bibinfo{year}{2002}).

\bibitem[{\citenamefont{Paul and Smith}(2004)}]{PSrev04}
\bibinfo{author}{\bibfnamefont{W.}~\bibnamefont{Paul}} \bibnamefont{and}
  \bibinfo{author}{\bibfnamefont{G.~D.} \bibnamefont{Smith}},
  \bibinfo{journal}{Rep. Prog. Phys.} \textbf{\bibinfo{volume}{67}},
  \bibinfo{pages}{1117} (\bibinfo{year}{2004}).

\end{thebibliography}

\end{document}